\documentstyle[prl,twocolumn,aps,epsf]{revtex}

\def\epsfig#1#2#3#4
         {
         \epsfysize=#2 \vbox{ \hglue#3 \epsfbox[#4]{#1} }
         }
\def\epsfigrot#1#2#3#4
         {
         \epsfxsize=#2
         \setbox\rotbox=\hbox to #2{\epsfbox[#4]{#1}}
         \vbox{\hglue#3 \rotl\rotbox}
         }
\newbox\rotbox
\input rotate
\begin{document}
\draft
\title{The Maxwell-Bloch Theory in Quantum Optics and the Kondo
Model}
\author{A. LeClair$^a$, F. Lesage$^b$, S. Lukyanov$^a$, H.
Saleur$^b$}
\address{$^a$ Newman Laboratory, Cornell University, Ithaca, NY
14853.}
\address{$^b$ Department of physics, University of Southern
California,
Los-Angeles, CA 90089-0484.}
\date{\today}
\maketitle

\begin{abstract}
In this letter, the problem of radiation in a fiber geometry
interacting with a two level atom is mapped onto the anisotropic
Kondo model.  Thermodynamical and dynamical properties are then
computed exploiting the integrability of this latter system.
We compute some correlation functions, decay rates and Lamb shifts. 
In turn this leads to an analysis of the classical limit
of the anisotropic Kondo model.
\end{abstract}
\vskip 0.2cm
\narrowtext
\def\dk{ \frac{dk}{\sqrt{2\pi}} }
%
%
\def\spht{
\centerline{Service de Physique Th\'eorique de Saclay}
\centerline{F-91191 Gif-sur-Yvette, France} }
%
%
\def\oti{{\otimes}}
\def\bra#1{{\langle #1 |  }}
\def\lb{ \left[ }
\def\rb{ \right]  }
\def\tilde{\widetilde}
\def\bar{\overline}
\def\hat{\widehat}
\def\*{\star}
\def\[{\left[}
\def\]{\right]}
\def\({\left(}		\def\BL{\Bigr(}
\def\){\right)}		\def\BR{\Bigr)}
	\def\BBL{\lb}
	\def\BBR{\rb}
%
%
\def\zb{{\bar{z} }}
\def\frac#1#2{{#1 \over #2}}
\def\inv#1{{1 \over #1}}
\def\half{{1 \over 2}}
\def\d{\partial}
\def\der#1{{\partial \over \partial #1}}
\def\dd#1#2{{\partial #1 \over \partial #2}}
\def\vev#1{\langle #1 \rangle}
\def\ket#1{ | #1 \rangle}
\def\rvac{\hbox{$\vert 0\rangle$}}
\def\lvac{\hbox{$\langle 0 \vert $}}
\def\2pi{\hbox{$2\pi i$}}
\def\e#1{{\rm e}^{^{\textstyle #1}}}
\def\grad#1{\,\nabla\!_{{#1}}\,}
\def\dsl{\raise.15ex\hbox{/}\kern-.57em\partial}
\def\Dsl{\,\raise.15ex\hbox{/}\mkern-.13.5mu D}
%
%
\def\th{\theta}		\def\Th{\Theta}
\def\ga{\gamma}		\def\Ga{\Gamma}
\def\be{\beta}
\def\al{\alpha}
\def\ep{\epsilon}
\def\vep{\varepsilon}
\def\la{\lambda}	\def\La{\Lambda}
\def\de{\delta}		\def\De{\Delta}
\def\om{\omega}		\def\Om{\Omega}
\def\sig{\sigma}	\def\Sig{\Sigma}
\def\vphi{\varphi}
%
%
\def\CA{{\cal A}}	\def\CB{{\cal B}}	\def\CC{{\cal C}}
\def\CD{{\cal D}}	\def\CE{{\cal E}}	\def\CF{{\cal F}}
\def\CG{{\cal G}}	\def\CH{{\cal H}}	\def\CI{{\cal J}}
\def\CJ{{\cal J}}	\def\CK{{\cal K}}	\def\CL{{\cal L}}
\def\CM{{\cal M}}	\def\CN{{\cal N}}	\def\CO{{\cal O}}
\def\CP{{\cal P}}	\def\CQ{{\cal Q}}	\def\CR{{\cal R}}
\def\CS{{\cal S}}	\def\CT{{\cal T}}	\def\CU{{\cal U}}
\def\CV{{\cal V}}	\def\CW{{\cal W}}	\def\CX{{\cal X}}
\def\CY{{\cal Y}}	\def\CZ{{\cal Z}}

\def\rvac{\hbox{$\vert 0\rangle$}}
\def\lvac{\hbox{$\langle 0 \vert $}}
\def\comm#1#2{ \BBL\ #1\ ,\ #2 \BBR }
\def\2pi{\hbox{$2\pi i$}}
\def\e#1{{\rm e}^{^{\textstyle #1}}}
\def\grad#1{\,\nabla\!_{{#1}}\,}
\def\dsl{\raise.15ex\hbox{/}\kern-.57em\partial}
\def\Dsl{\,\raise.15ex\hbox{/}\mkern-.13.5mu D}
%
%
%
\font\numbers=cmss12
\font\upright=cmu10 scaled\magstep1
\def\stroke{\vrule height8pt width0.4pt depth-0.1pt}
\def\topfleck{\vrule height8pt width0.5pt depth-5.9pt}
\def\botfleck{\vrule height2pt width0.5pt depth0.1pt}
\def\Zmath{\vcenter{\hbox{\numbers\rlap{\rlap{Z}\kern
0.8pt\topfleck}\kern 2.2pt
                   \rlap Z\kern 6pt\botfleck\kern 1pt}}}
\def\Qmath{\vcenter{\hbox{\upright\rlap{\rlap{Q}\kern
                   3.8pt\stroke}\phantom{Q}}}}
\def\Nmath{\vcenter{\hbox{\upright\rlap{I}\kern 1.7pt N}}}
\def\Cmath{\vcenter{\hbox{\upright\rlap{\rlap{C}\kern
                   3.8pt\stroke}\phantom{C}}}}
\def\Rmath{\vcenter{\hbox{\upright\rlap{I}\kern 1.7pt R}}}
\def\Z{\ifmmode\Zmath\else$\Zmath$\fi}
\def\Q{\ifmmode\Qmath\else$\Qmath$\fi}
\def\N{\ifmmode\Nmath\else$\Nmath$\fi}
\def\C{\ifmmode\Cmath\else$\Cmath$\fi}
\def\R{\ifmmode\Rmath\else$\Rmath$\fi}

\section{Introduction}

The interaction of radiation with two-level atoms is a fundamental
problem with many important applications.
The model goes under the name of the Maxwell-Bloch theory in
quantum optics.
The problem has been extensively studied theoretically
in a framework where the radiation is essentially classical
and the so-called slowly varying envelope and rotating wave
approximations are commonly made, which are valid near
resonance\cite{reson}.
In one dimension, fully quantum integrable versions of this system,
refered to
as the reduced Maxwell-Bloch theory, have been solved
exactly\cite{rupasov}.
In some applications, off-resonance effects are likely to be
significant;
for example in doped photonic band gap materials.
In this letter, we show that the single atom case is exactly solvable
without making any near-resonance  approximations, by mapping the
problem onto
the anisotropic Kondo model.  Furthermore, we are able to  compute
important physical quantities that have hitherto been inaccessible in
previous treatments of even the approximate reduced Maxwell-Bloch
theory.  In particular, we  compute
thermodynamic
properties in the form of impurity corrections to the
Stephan-Boltzman law,
and some electric field correlators.
From the latter we obtain the atomic spontaneous emission decay rate and 
the Lamb shifted energy splitting. 

\section{The models}

The interaction of radiation with
a two-level atomic impurity at $x_0$  is described by the total
hamiltonian
\begin{equation}
\label{hamilo}
H=H^{atom}+H^{field}+H^{int}
\end{equation}
with
\begin{eqnarray}
H^{field}&=&\frac{1}{2} \int_{-\infty}^\infty
dx [(\partial_t \phi)^2+(\partial_x\phi)^2] ,
\\ \nonumber
H^{atom}&=&\frac{ \omega_0}{2} \sigma_3 ,
{}~~~~~H^{int}=\frac{\beta}{2} \partial_t \phi (x_0) \left( \sigma_++
\sigma_- \right),
\label{hamili}
\end{eqnarray}
and the conventions  $[ \sigma_3 , \sigma_\pm ] = \pm 2 \sigma_\pm $,
$[\sigma_+ , \sigma_- ] = \sigma_3 $. We have made a reduction from
the Maxwell theory to a one-dimensional
fiber geometry, following the conventions in \cite{lec}. The
dimensionless coupling constant $\beta$ is determined by the
strength $d$ of the dipole transition and the effective
cross-sectional
area of the fiber,
$\beta=\sqrt{\frac{16\pi}{ {\cal A}_{eff}}} d$.
The coupling $\beta$ is also related to the lifetime $\tau$
of the excited state of the atom;  to lowest order,
$\gamma = 1/\tau  = \beta^2 \omega_0 /4 $.

The hamiltonian (\ref{hamili}) is known in the condensed matter
literature as the spin boson hamiltonian \cite{GHM},\cite{Leg}.  It
can be mapped onto the anisotropic Kondo hamiltonian $H_K$: defining
the unitary operator
\begin{equation}
\label{uu}
U = \inv{\sqrt{2}}
\exp \( i \frac{\beta}{2} \sigma_3 \phi (x_0) \) (\sigma_3
+ \sigma_+ + \sigma_- ),
\end{equation}
then  $H=U^\dagger H_K U$, where
\begin{equation}
H_K =H^{field}+\frac{ \omega_0}{2} (\sigma_+ e^{i\beta \phi(x_0)}+
\sigma_- e^{-i\beta \phi(x_0)}).
\end{equation}
This allows us to formulate the quantum mechanics in either the
``optical picture" based on $H$ or the ``Kondo picture" based on
$H_K$.


In order to complete the map to the conventional
Kondo model on the half-line $x<0$,
we
fold the system.  Let $\phi_L (x+t),
\phi_R (x-t)$ be the left and right-moving
components  in the bulk, with
$\phi = \phi_L + \phi_R$.
Let us set $x_0 = 0$, and
define even and odd fields
in the region $x<0$
\begin{eqnarray}
\varphi^e_L (x,t) &=& \beta \left( \phi_L (x,t) + \phi_R (-x,t)
\right)
\nonumber\\
\varphi^o_L (x,t) &=& \beta \left(  \phi_L (x,t) - \phi_R (-x,t)
\right)
\nonumber \\
\varphi^e_R (x,t) &=& \varphi^e_L (-x,t),
{}~~~~~
\varphi^o_R (x,t) = - \varphi^o_L (-x,t).
\end{eqnarray}
Introduce new fields $\phi^{e,o}=\varphi^{e,o}_L+\varphi^{e,o}_R$.
The odd and even fields decouple; the odd field just becomes
a free field, while for the even field  the hamiltonian reads
\begin{eqnarray}
\label{hamil}
H_K^e  =   \frac{1}{2}\int_{-\infty}^0 dx \
[8\pi g \Pi^{2,e}+\frac{1}{8\pi g}  (\partial_x\phi^e)^2]
\nonumber \\
+ \frac{\omega_0}{2}  (\sigma_+e^{i\phi(0)^e/2}+\sigma_-
e^{-i\phi^e(0)/2}),
\end{eqnarray}
where $g=\beta^2 /4\pi$.

Recent progress in the understanding of the anisotropic Kondo problem
will now allow us to compute exactly the quantities of interest in
the optics problem.

\section{Thermodynamics}

We start with thermodynamics. Consider the system at finite
temperature $T$.
The partition function of
the optical system is identical to the Kondo one:
$Z = Tr e^{-H/T} = Tr e^{-H_K /T} $.
Without the atom,
the photon energy spectrum is  given by the usual Plank distribution.
The total partition function is
\begin{eqnarray}
\label{zfree}
\log Z |_{g=0}  &=& - \frac{L}{2\pi} \int_{-\infty}^\infty dk
\log ( 1 - e^{-|k|/T} ) + \log Z_{\rm atom}
\nonumber
\\
&=& \frac{\pi}{6} LT + \log \( 2 \cosh\( \frac{\omega_0}{2T} \) \),
\end{eqnarray}
where $L$ is the length of the system.
%
%
%
Of particular  interest is the mean energy density
\begin{equation}
\label{derpart}
U(T) = \frac{T^2}{L}  \d_T \log Z
= \frac{\pi}{6} T^2 - \frac{\omega_0}{2L} \tanh \(
\frac{\omega_0}{2T} \).
\end{equation}
The first term in (\ref{derpart}) is the 1d analog of the
Stephan-Boltzman law.

In the interacting theory, we work with
equation (\ref{derpart}).
The partition function
factorizes into a bulk term and an impurity contribution
$Z = Z_{\rm bulk} \cdot Z_{\rm imp}$.
The bulk contribution, including both $\phi^{e,o}$,
is the same as
(\ref{zfree}):
$\log Z_{\rm bulk} = \frac{\pi}{6} LT$.

The impurity contribution in the context of the Kondo problem has
recently been studied using novel techniques in
\cite{baza}\cite{unif}. In particular, the partition function
$Z_{\rm imp}$ is the trace of the monodromy
matrix  studied in detail in \cite{baza} \footnote{For the
conventions in \cite{baza} ,  the
`spectral parameter' argument $\lambda$ of the
monodromy matrix  $T(\lambda)$ is a function of
$\omega_0$ and $T$, namely $\lambda = \omega_0  (2\pi T)^{g-1}
/2$.
The parameter $\omega_0$ has anomalous scaling dimension $-g$,
which is consistent with the 
the renormalization group equation
derived in the
optical picture in \cite{lec}.}.

It is convenient to express physical quantities in terms of
a ``Kondo temperature"  $T_K$, defined such that the leading
low temperature behaviour is $\log Z_{\rm imp} \sim T_K/T$.
The Kondo temperature is then  a fixed function of the energy
splitting and $g$
\begin{equation}
\label{kondoT}
T_K = \inv{\sqrt{\pi}}
\frac{ \Gamma \( \frac{1-2g}{2-2g} \) }{\Gamma \( \frac{2-3g}{2-2g}
\) }
\( \frac{\omega_0}{2} \Gamma (1-g) \)^{\inv{1-g}} .
\end{equation}
The leading high and low temperature behaviors for arbitrary $g$
are then  as follows.
At low temperatures,
\begin{equation}
\label{low}
\log Z_{\rm imp} \asymp \frac{T_K}{T} + \sum_{n=0}^\infty
a_n (g) ~ \( \frac{T}{T_K} \)^{2n+1}
\end{equation}
where  $a_0 = \frac{\pi}{6} \tan \frac{\pi g}{2-2g} $,
\begin{equation}
a_1 =
\frac{\pi^2 (1-4g)(4-g)}{360}
\frac{ \Gamma^3 \( \frac{1-2g}{2-2g} \) \Gamma \( \frac{3}{2-2g} \) }
{ \Gamma^3 \( \frac{2-3g}{2-2g} \) \Gamma \( \frac{3g}{2-2g} \) }
{}.
\end{equation}
Higher $a_n (g) $, up to $n=3$, can be deduced from  results in
\cite{baza}.
At high temperatures, one finds
\begin{equation}
\label{high}
Z_{\rm imp} = 2 +
\tilde{a}(g) \( \frac{T_K}{T} \)^{2-2g} + \CO \( \frac{T_K}{T}
\)^{4-4g}
\end{equation}
where
\begin{equation}
\tilde{a} (g) = 4\pi^2
\frac{ \Gamma(1-2g)}{\Gamma^4 (1-g) }
\(
\inv{2\sqrt{\pi}}
\frac{ \Gamma \( \frac{2-3g}{2-2g} \) }{\Gamma \( \frac{1-2g}{2-2g}
\) }
\)^{2-2g}
\end{equation}

The mean energy density $U(T)$ has the form
$U = U_{\rm bulk} + U_{\rm imp}$ where $U_{\rm bulk} = \pi T^2 /6 $.
At  low  and high temperatures for arbitrary $g$, one finds
\begin{eqnarray}
U_{\rm imp} &\approx&
\inv{L}  \( - T_K + a_0 ~ \frac{T^2}{T_K} \)   ,~~~~~~~~~      T \ll
T_K
,
\nonumber
\\
&\approx&
- \tilde{a}  (1-g) \frac{T}{L} \( \frac{T_K}{T} \)^{2-2g} ,  ~~~~~T
\gg T_K .
\end{eqnarray}

In the quantum optical context, $g$ is generally very small, thus it
is more relevant to expand $Z_{\rm imp}$ in powers of $g$.
For small $g$,  $T_K \approx \omega_0/2$.
In \cite{baza}, the low temperature expansion of $Z_{\rm imp}$ is
expressed
in terms of the so-called local integrals of motion $I_{2n-1}$.
With the conventions of \cite{baza}, one can show that
\begin{equation}
g^{n-1} I_{2n-1} (g) \vert_{g=0} = - \inv{2^{n+1}}
\frac{ (n-1)!}{(2n-3)!!}  ~ | B_{2n} |,
\end{equation}
where $B_{2n}$ are the Bernoulli numbers.  Using this, one can
show that to order $g$ the partition function at arbitrary
temperature
is given by
\begin{eqnarray}
\label{allT}
Z_{\rm imp} &=& 2 \cosh \( \frac{T_K}{T} \)
+ 2  g \frac{T_K}{T}  \sinh \( \frac{T_K}{T} \) \times
\nonumber
\\
&~& ~~~~~~ Re \[ \Psi \( 1 + \frac{i T_K}{\pi T} \)
- \log\( \frac{T_K}{\pi T} \) \],
\end{eqnarray}
where
$\Psi (x) = \d_x \log \Gamma (x)$.
One can easily check that (\ref{allT})
correctly reproduces the high and low temperature expansions
in (\ref{low}) (\ref{high}) at small $g$.

The classical limit of the sine-Gordon model thermodynamics  had been
partially studied in \cite{F}. We see here that  there are, in fact,
physical probes  exploring the
different Bethe ansatz excitations: these are  the spin $j$
impurities, whose thermal properties depend on the pseudoenergies
$\epsilon_{2j}$ of the  thermodynamic bethe ansatz. For instance,
when $g=0$,
$Z_{\rm imp}$ corresponds to the partition function of a
decoupled two-level system with energies $\pm \omega_0 /2$:
$Z_{\rm imp} = 2 \cosh (\omega_0 /2T)$, and this can be recovered
using the expressions for $\epsilon_1$ in \cite{F}.

The mean energy density can be easily obtained for all $T$
from (\ref{derpart}) and (\ref{allT}).


\section{Photon Correlation Functions}

Let $E$ denote the relevant component of the electric field
perpendicular
to the fiber.
Given some initial state $|i\rangle$, or some mixed state with
density matrix $\rho = \sum_i P_i |i\rangle \langle i|$, one is
interested in
the intensity of the radiation in the electric field
\begin{equation}
\label{intense}
I(x) = \inv{4\pi}  \langle  E^{(-)} (x,t) E^{(+)} (x,t)
\rangle_\rho,
\end{equation}
where $\langle \CO \rangle_\rho = Tr \rho \CO$,
and $E^{(-)}$ ($E^{(+)}$) is the creation (annihilation) part of the
field, $E = E^{(+)} + E^{(-)}$.
In the quantum measurement theory of Glauber, $I(x)$
is proportional to the number of photons detected at $x$ given
the initial state specified by $\rho$.  One is also interested
in the power spectrum $S(\omega )$ of detected photons 
satisfying $\int_0^\infty d\omega S(\omega) = I(x)$.
One can express the above quantities in terms of the usual electric
field correlators.  Taking as the definition of
$E^{(\pm)}$:
\begin{equation}
E^{(\pm)} (x,t) = \pm \frac{i}{2\pi}
\int_{-\infty}^\infty dt' \inv{t' - t \pm i \epsilon} E (x,t'),
\end{equation}
one can show that if $|\psi\rangle$ is an eigenstate of the
hamiltonian
with energy $E_\psi$, then
$\langle \psi' | E^{(+)} |\psi \rangle
= \langle \psi' | E |\psi \rangle $ if $E_{\psi'} \leq E_\psi $
and zero otherwise.  Likewise,
$\langle \psi' | E^{(-)} |\psi \rangle
= \langle \psi' | E |\psi \rangle $ if $E_{\psi'} \geq E_\psi $
and zero otherwise.   Inserting complete sets of states in
(\ref{intense}), one finds
\begin{equation}
\label{Power}
S(\omega) = \theta (\omega) \inv{8\pi^2}  \int_{-\infty}^\infty d\tau
e^{-i \omega \tau}
\langle E (x, \tau) E (x,0) \rangle_\rho .
\end{equation}

The power spectrum $S(\omega)$  is non-zero only for density matrices
$\rho$
containing excited states, e.g. a thermal distribution, since the
integral in (\ref{Power}) is proportional to $\theta (-\omega)$
for $\rho = |0\rangle \langle 0|$.  For example, 
such correlators were computed for the harmonic oscillator version
of the present problem in \cite{konik}. In this paper, in order to
illustrate our approach to correlation functions, we deal with
the simpler quantity $\langle 0| E(x,t) E(x,0) |0\rangle  $,
which probes how easy it is to {\it add} a photon to (rather
than {\it detect} a photon in)  the vacuum.
In the reduction to one dimension,
$E  = \sqrt{ \frac{4\pi}{\CA_{eff}} } \d_t \phi$.
Thus we define
\begin{equation}
\label{power}
i(x) =    \langle 0|
: \d_t \phi  (x,t) \d_t \phi (x,t) :  |0 \rangle = \int_0^\infty
d\omega ~ s(\omega) .
\end{equation}
We define the normal ordering in (\ref{power})
such that the free photon  vacuum energy $|\omega|/4\pi$ is
subtracted from $s(\omega)$.
To lowest order in perturbation theory in the optical picture,
 one has
\begin{equation}
\label{pert}
s(\omega)=i \frac{g\omega_0}{2} {\omega^2\over \omega^2-\omega_0^2+i
\epsilon} e^{2i \vert \omega x\vert}.
\end{equation}

All the previous quantities were defined in the unrotated and
unfolded theory.
The rotation by the unitary operator $U$ changes
the operator  $\partial_t  \phi$ to
\begin{equation}
\partial_t\phi (x,t)\rightarrow \partial_t \phi(x,t)-\frac{\beta}{2}
\sigma_3 \delta(x) .
\end{equation}
Since the detector is
taken to be away from the impurity (and at $x<0$ in our
conventions), the second term of the
previous expression can be dropped.  Then under folding,
we find
\begin{equation}
\partial_t \phi(x,t) =  \frac{1}{2\beta} \( \partial_t \phi^e(x,t)+
 \partial_t \phi^o(x,t) \) ,
\end{equation}
where now the odd and even field are defined on the negative
axis.  The odd field is free and has Dirichlet boundary
conditions,  $\phi^o(0)=0$, whereas
the even field has Neumann boundary conditions
and interacts with the boundary.  When computing $s(\omega )$
the odd field will not contribute because the Wick contraction
is substracted.  Thus
\begin{equation}
i(x)=\frac{1}{16 \pi g} \langle 0| :\partial_t \phi^e (x,t)^2:
|0\rangle .
\end{equation}

In order to compute these correlation functions, we will make use
of the techniques used for the anisotropic Kondo model in
\cite{lss}.
There, the ``natural" basis of the excitations of the system,
as described by the massless sine-Gordon model, is used
to compute correlation functions.
At first sight, the problem appears untractable since, as
$g\rightarrow 0$, the  spectrum of the sine-Gordon theory
consists of a very large number of particles: soliton, anti-soliton
and
$n$ bound states or
breathers where $n\simeq 1/g$.  The problem simplifies
however when  considering  the different contributions  of
these particles to the
correlators.  A quick way to see this
is to consider 
$<\partial_z\phi(z,\bar{z}) \partial_{z'}\phi (z',\bar{z}')
>={2g}/{(z-z')^2}$,
which is  unaffected by the boundary interaction. 
(Above, $z=x+iy$, $\bar{z}=x-iy$, and $y=it$)  
This correlator
can also be computed using form factors, and expands as a sum of
positive contributions. This gives a good idea of the order of
magnitude of neglected terms when only the most important excitations
are taken into account: for instance  the one breather  is exact to
order $g$ and
is an excellent approximation for values of $g$ up to $g \sim 1/5$
(see below).

The general expression for the current-current
correlator is
\begin{eqnarray}
<\partial_{\bar{z}}\phi(z,\bar{z}) & &\partial_{z'}\phi
(z',\bar{z}')>=
\sum_{n=0}^\infty \frac{1}{n!} \sum_{\{\epsilon_i\}} \int
\prod_{i=1}^n \frac{d\theta_i}{2\pi}  \\  \nonumber
 R(\theta_i-\theta_B)\cdot & &
\vert f^z(\theta_1,...,\theta_n)\vert^2 \
e^{-(\bar{z}+z')(M_{\epsilon_1}
 e^{\theta_1}+...+M_{\epsilon_n}e^{\theta_n})}
\end{eqnarray}
where $f^z$ is the form factor of the right moving current operator
and
the dot product means that there is a contraction between the
isotopic indices of the $R$ matrices and that of the form factors.

\def\omh{{\hat{\omega}}}
The one breather has mass $M_1=2 \sin(\frac{\pi g}{2-2g})$ and using
the
results of \cite{lss} we find the intensity to be
\begin{equation}
i(x)=- \int_0^{\infty} \frac{d\omega}{4\pi}
 \omega  \[ K_1 ( \log ( \frac{\omega}{M_1 T_B} ) )-1 \] 
 e^{-2 \omega  x}
\end{equation}
where for convenience we have defined $T_B=T_K/\tan(\pi g/2(1-g))$,
and
$K_1$ is related to the reflection matrix for the breather
by $K_1(\theta)=R_1(i\frac{\pi}{2}-\theta)$, where
\begin{equation}
\label{reflec}
R_1(\theta)={\tanh(\frac{\theta}{2}-i\frac{\pi}{4}
(\frac{g}{(1-g)})
\over
\tanh(\frac{\theta}{2}+i\frac{\pi}{4}(\frac{g}{(1-g)})}.
\end{equation}
The `power spectrum'   $s(\omega)$ thus  takes a
very simple form
\begin{equation}
s(\omega)=
i \frac{\gamma}{2\pi} \( \frac{\omega^2}{\omega^2 - \omh_0^2 + i \omega \gamma}
\) e^{-2 i \omega x},
\end{equation}
where 
\begin{equation}
\omh_0 = 2 \cos \( \frac{\pi g}{2-2g} \)  T_K  , ~~~~~
\gamma = 2 \sin \( \frac{\pi g}{1-g} \) T_K . 
\end{equation}
This result reproduces  the perturbation theory computation
at small $g$ (\ref{pert}).
Near resonance $\omega \approx \omh_0$, $s(\omega)$ takes on the
familiar Wigner-Weisskopf Lorentzian lineshape, with 
a Lamb shifted  resonant frequency $\omh_0$ and a 
half-width $\gamma/2$:  
\begin{equation}
s(\omega) = \frac{\gamma^2}{8\pi} \( \frac{\omh_0}{ (\omega - \omh_0)^2 
+ (\gamma/2)^2 } \) e^{-2 i \omega x} .   
\end{equation}
We remark that since $\gamma$ is twice the imaginary part of
the pole of $R_1$  in the momentum variable $k=M_1 T_B e^\theta$, 
in a sense $\gamma$ represents the exact decay rate into the first 
breather.  In the small $g$ limit, the first breather is identified with the
photon and thus $\gamma$ agrees with the lowest order perturbative
result $\gamma = \pi g \omega_0$. 

One can check that this one-breather approximation is better than the
latter perturbative result.
Actually, it turns out that it is an excellent approximation for
the correlators for values of $g$ at least up to $g=1/5$!
Consider for instance the case  $g=1/5$.
Writing
$s(\omega)=\frac{\omega}{4\pi} e^{-2 i \omega x} F(\omega/T_B)$,
where $F$ is a universal function,  let us compare the
contributions to the imaginary part of
$F$ coming from other intermediate states.  (The contributions
for the real part have similar properties).
This can be done by using explicit  formulas for  form
factors deduced from \cite{S}.
We display  in figure 1 the next dominant
contributions, i.e.  the one particle form factor of the breather
$n=3$ and the two particle soliton/anti-soliton form factor.
It is clear from the figure that these contributions, which
can barely be seen on the graph, are
negligible when compared to the one breather contribution
for most of the range.  In the deep IR, when $\omega/T_B>10$,
then the soliton/anti-soliton form factor starts to contribute
(indeed it is necessary to get the correct IR exponent)
but the function $F$ is very small in that limit
and this is certainly not
the dominant region.  The perturbative expression
is also given on the graph to give an idea of the differences.
The $g$ dependence of the correlator is then entirely
included in the reflection matrix of the breather
(\ref{reflec}) and
is very easy to extract from the previous formulae.
\vbox{
\epsfysize=8cm
\epsfxsize=8cm
\epsffile{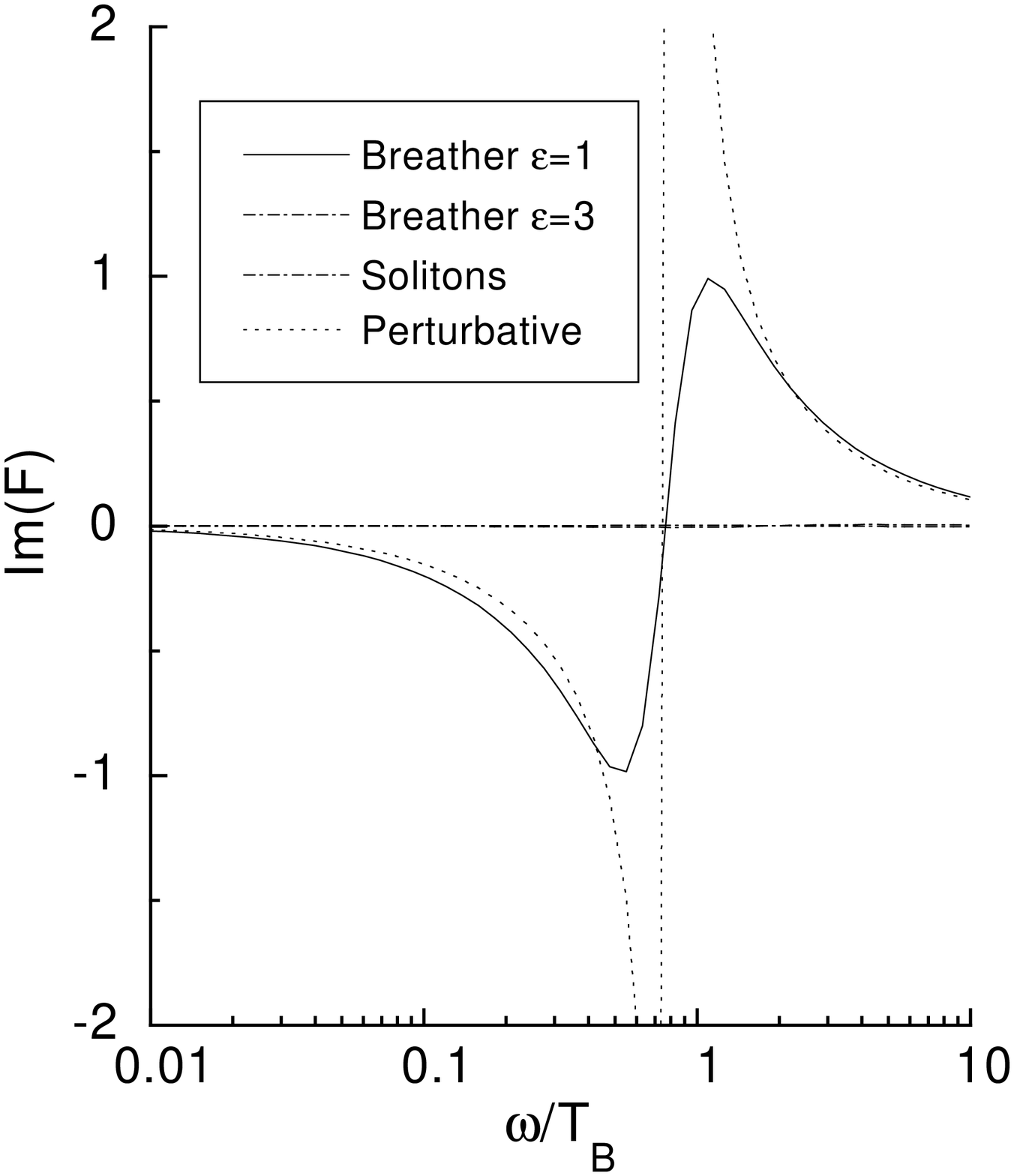}
\begin{figure}
\caption[]{\label{fig1} Accuracy comparison for various
form factor contributions for $g=1/5$.}
\end{figure}}



\begin{references}

\bibitem{reson} L. Allen and J. H. Eberly, {\it Optical Resonance
and Two-Level Atoms}, Dover Publications, New York, 1987.

\bibitem{rupasov} V. I. Rupasov, JETP Lett. 36 (1982) 142.

\bibitem{lec} A. LeClair, {\it QED for a Fibrillar Medium of
Two-Level
Atoms}, hep-th/9604100, to appear in Phys. Rev. A.

\bibitem{GHM} F. Guinea, V. Hakim, A. Muramatsu, Phys. Rev. B32
(1985) 4410.

\bibitem{Leg}A. J. Leggett, S. Chakravary, A. T. Dorsey, M. P.
Fisher, A. Garg, W. Swerger, Rev. Mod. Phys. 59 (1987) 1.

\bibitem{baza} V. V. Bazanov, S. L. Lukyanov, and A. B.
Zamolodchikov,
Commun. Math. Phys. 177 (1996), 381.

\bibitem{unif} P. Fendley, F. Lesage, H. Saleur,
to appear  in J. Stat. Phys.; cond-mat/9510055.

\bibitem{F} M. Fowler, Phys. Rev. B26 (1982) 2514.

\bibitem{lss} F. Lesage, H. Saleur, S. Skorik, Phys. Rev. Lett. 76
(1996) 3388, cond-mat/9512087; Nucl. Phys. {\bf B}474, (1996) 602,
cond-mat/9603043.

\bibitem{S} F. Smirnov, ``Form factors in completely integrable
models
of quantum field theory'', World Scientific.

\bibitem{konik}  R. Konik and A. LeClair, {\it Scattering Theory
of Oscillator Defects in an Optical Fiber}, CLNS 96/1442. 

\end{references}
\end{document}